\documentstyle[11pt]{article}
\begin{document}
\newcommand{\Si}{\Sigma}
\newcommand{\tr}{{\rm tr}}
\newcommand{\ad}{{\rm ad}}
\newcommand{\Ad}{{\rm Ad}}
\newcommand{\ti}[1]{\tilde{#1}}
\newcommand{\om}{\omega}
\newcommand{\Om}{\Omega}
\newcommand{\de}{\delta}
\newcommand{\al}{\alpha}
\newcommand{\te}{\theta}
\newcommand{\vth}{\vartheta}
\newcommand{\be}{\beta}
\newcommand{\la}{\lambda}
\newcommand{\La}{\Lambda}
\newcommand{\D}{\Delta}
\newcommand{\ve}{\varepsilon}
\newcommand{\ep}{\epsilon}
\newcommand{\vf}{\varphi}
\newcommand{\G}{\Gamma}
\newcommand{\ka}{\kappa}
\newcommand{\ip}{\upsilon}
\newcommand{\Ip}{\Upsilon}
\newcommand{\ga}{\gamma}
\newcommand{\ze}{\zeta}
\newcommand{\si}{\sigma}
\def\vka{\varkappa}
\def\SR{\Sigma_{g,n}}
\def\bfa{{\bf a}}
\def\bfb{{\bf b}}
\def\bfc{{\bf c}}
\def\bfd{{\bf d}}
\def\bfm{{\bf m}}
\def\bfn{{\bf n}}
\def\bfp{{\bf p}}
\def\bfu{{\bf u}}
\def\bfv{{\bf v}}
\def\bft{{\bf t}}
\def\bfx{{\bf x}}
\newcommand{\li}{\lim_{n\rightarrow \infty}}
\newcommand{\mat}[4]{\left(\begin{array}{cc}{#1}&{#2}\\{#3}&{#4}
\end{array}\right)}
\newcommand{\mathr}[9]{\left(\begin{array}{ccc}{#1}&{#2}&{#3}
\\{#4}&{#5}&{#6}\\
{#7}&{#8}&{#9}
\end{array}\right)}
\newcommand{\beq}[1]{\begin{equation}\label{#1}}
\newcommand{\eq}{\end{equation}}
\newcommand{\beqn}[1]{\begin{eqnarray}\label{#1}}
\newcommand{\eqn}{\end{eqnarray}}
\newcommand{\p}{\partial}
\newcommand{\di}{{\rm diag}}
\newcommand{\oh}{\frac{1}{2}}
\newcommand{\su}{{\bf su_2}}
\newcommand{\uo}{{\bf u_1}}
\newcommand{\GL}[1]{{\rm GL}({#1},{\bf C})}
\newcommand{\SL}[1]{{\rm SL}({#1},{\bf C})}
\newcommand{\sll}[1]{{\rm sl}({#1},{\bf C})}
\def\sln{{\rm sl}(N,{\bf C})}
\newcommand{\gl}{gl(N,{\bf C})}
\newcommand{\PSL}{{\rm PSL}_2({\bf Z})}
\def\f1#1{\frac{1}{#1}}
\newcommand{\rar}{\rightarrow}
\newcommand{\upar}{\uparrow}
\newcommand{\sm}{\setminus}
\newcommand{\ms}{\mapsto}
\newcommand{\bp}{\bar{\partial}}
\newcommand{\bz}{\bar{z}}
\newcommand{\bA}{\bar{A}}
\newcommand{\bL}{\bar{L}}
\newcommand{\bAp}{\bar{A}^\prime}
\newcommand{\begarl}{\begin{array}{l}}
\newcommand{\enarr}{\end{array}}
\newcommand{\begarc}{\begin{array}{c}}
\newcommand{\sect}[1]{\setcounter{equation}{0}\section{#1}}
\renewcommand{\theequation}{\thesection.\arabic{equation}}
\newtheorem{predl}{Proposition}[section]
\newtheorem{defi}{Definition}[section]
\newtheorem{rem}{Remark}[section]
\newtheorem{cor}{Corollary}[section]
\newtheorem{lem}{Lemma}[section]
\newtheorem{theor}{Theorem}[section]

\vspace{0.3in}
\begin{flushright}
 ITEP-TH-60/00\\
\end{flushright}
\vspace{10mm}
\begin{center}

{\Large\bf
$W$-geometry and Isomonodromic Deformations}
\footnote
{Talk given at the CRM Workshop on
Isomonodromic Deformations and Applications in Physics, Montreal, Canada, May 1-6, 2000}
\footnote{The work is supported in part by the grants INTAS 99-01782,
RFFI-00-02-16530 and 96-15-96455 for support of scientific schools. }
\\
\vspace{5mm}
M.A.Olshanetsky
\\
{\sf Institute of Theoretical and Experimental Physics, Moscow, Russia,}\\
{\em e-mail olshanet@heron.itep.ru}\\

\vspace{5mm}
\end{center}

\begin{abstract}
We introduce new times in the monodromy preserving equations.
While the usual times related to the moduli of complex structures of
Riemann curves such as coordinates of marked points, we
consider the moduli of generalized complex structures ($W$-structures).
We consider linear differential matrix equations depending on $W$-structures
on an arbitrary Riemann curve. The monodromy preserving equations have
a Hamiltonian form. They are derived via the symplectic reduction
procedure from a free gauge theory as well as the associate linear problems.
The quasi-classical limit of isomonodromy problem leads to  integrable
hierarchies of the Hitchin type. In this way the generalized complex structures
parametrized the moduli of these hierarchies.
\end{abstract}
\vspace{0.3in}

\bigskip

\section {Introduction}
\setcounter{equation}{0}

The monodromy preserving equations arise from the consideration of
a linear matrix equation on ${\bf C}P^1$
\beq{1}
(\p + A)\Psi=0.
\eq
Assume for simplicity that meromorphic matrix function $A$ has only the first
order poles
$$
A=\sum_{a=1}^n\frac{p_a}{z-x_a}.
$$
The independence of the monodromies of $\Psi$ around the poles on positions of the
poles $x_a$  gives rise to the monodromy preserving equations. In this case
 monodromy preserving equations  is the Schlesinger system.
This system is nontrivial if the number
of poles $n>3$. In particular, for $n=4$ and for the two by two matrices
the Schelesinger system is equivalent to the famous Painl\`{e}ve VI equation.
All these equations can be
described as non autonomous Hamiltonian systems, where the role of times
plays by the positions of poles. The positions of the poles correspond
to some fixed complex structure on a sphere ${\bf C}P^1(n)$ with $n$ the marked points.
 On the other hand the complex structure
can be encoded in the Beltrani differential $\mu$. It allows to
deform the operator of comlex structure $\bp\rar \bp+\mu\p$. The Beltrami
differentials are defined up to  diffeomorphisms of ${\bf C}P^1(n)$.
Locally $\mu$ can be considered as an
 element of the Teichm\"{u}ller space $H^1({\bf C}P^1(n),\G)$,
$(\dim H^1({\bf C}P^1(n),\G)=n-3)$. Here
$\G$ is the space of smooth sections of the tangent bundle $T{\bf C}P^1$.
For example, the Beltrami differential
 can be chosen in the form
$$
\mu=\sum_{a=1}^n(x_a-x^0_a)\bp\chi_a(z,\bz),
$$
where $\chi_a(z,\bz)$ is a characteristic function of a small neighborhood
of the fixed point $x_a^0$ (see (\ref{chi})). In this way the Beltrami differentials
are closely related to the geometry of ${\bf C}P^1(n)$.

There exists a   generalization of this geometry that comes from the higher order
differentials $\mu_j\in H^1({\bf C}P^1(n),\G^{\otimes j})$. Roughly speaking this
generalization allows to deform the operator $\bp$ as
$$
\bp+\sum_j\mu_j\p^j.
$$
The hidden geometry behind these deformations is known as $W$-geometry
\cite{P}.

In this talk I present an attempt to introduce the  times connected
with $\mu_j,~j>1$ in the Isomonodromic deformation problem in the similar
way as it was done for the Beltrami differentials $\mu\sim\mu_1$.
We start from the flat $\SL{N}$-bundles over a Riemann curve  $\Si_{g,n}$
of genus $g$ with $n$ marked points. It is a phase space for the
standard Isomonodromic deformation problem on $\Si_{g,n}$ \cite{LO1}.

Then we consider the generalized deformations of complex structures on $\Si_{g,n}$
and the associated Isomonodromic deformation problems. The
symplectic reduction procedure leads to meaningful equations and allows
to write down the relevant linear systems.

The next point discussed here is the quasi-classical limit from the
monodromy preserving equations to integrable hierarchies. It means in particular
that in  (\ref{1}) the operator $\p$ is replaced by $\ka\p$ and $\ka\to 0$. In this case
one comes from the Schlesinger system to the Gaudin system, where
positions of the points are merely parameters \cite{Ta}.
The approach presented in \cite{Ta} can be generalized on the Isomonodromic
deformations with higher times.
We can fix the higher order differentials up to the order $N-1$.
In this case the classical limit leads to the hierarchies of completely
integrable hierarchies. When the $W$-moduli vanish they pass into
the Hitchin systems \cite{Hi,Ne}. In this way  the $W$-moduli can be
considered as deformation parameters of integrable hierarchies.

\section {Isomonodromy preserving equations as reduced Hamiltonian systems}

\setcounter{equation}{0}

{\bf 1. The upstairs phase space ${\cal R}^N$.}
 We consider a stable bundle $\SL{N}$ bundle $E$ over a Riemann curve
$\Si_{g,n}$ of genus $g$ with $n$ marked points.
The upstairs phase space ${\cal R}^N$ is the collection
of the coadjoint generic orbits ${\cal O}_a$ of $\SL{N}$ in the marked points
$x_a,~a=1,\ldots,n$
along with the  connections
$$
\nabla^{(1,0)}:~\Om^0(\Si_{g,n},E)\to\Om^{(1,0)}(\Si_{g,n},E),~~
\nabla^{(0,1)}:~\Om^0(\Si_{g,n},E)\to\Om^{(0,1)}(\Si_{g,n},E).
$$
Over a small disk  $\nabla^{(1,0)}=\ka\p+A$, $\nabla^{(0,1)}=\bp+\bAp$
and
\beq{2.1}
{\cal R}^N=\{\ka\p+A,\bp+\bAp;({\cal O}_1,\ldots,{\cal O}_n)\}.
\eq
Here $\ka$ is the level. It will play the role of the Plank constant later,
when we shall consider integrable theories arising in the classical limit $(\ka\to 0)$.
The connections can have first order poles at the marked points.

Represent  $p_a\in{\cal O}_a$  as
$p_a=g_ap_a^0g_a^{-1}$,
$g_a\in\SL{N}$ and $p_a^0$ fixes the conjugate class of ${\cal O}_a$.
The Kirillov-Kostant symplectic form on ${\cal O}_a$  in this parameterizations is
$$
\tr (d(p_ag_a^{-1})~dg_a)=<d(p_ag_a^{-1})~dg_a>.
$$
The symplectic form on the whole space ${\cal R}^N$ is
\beq{2.2}
\om=\int_{\Si_{g,n}}<dA~ d\bAp>+\om_{\cal O} ,~~~
\om_{\cal O}=\sum_{a=1}^n<d(p_ag_a^{-1})~ dg_a>
\eq

The form is invariant under the action of the gauge group
${\cal G}_N =\{f\}\sim\Om^0(\Si_{g},{\rm Aut}E)$
\beq{2.3}
A\rar f^{-1}\ka\p f+f^{-1}Af,~~\bAp\rar f^{-1}\bp f+f^{-1}\bAp f,
\eq
\beq{2.3a}
p_a\rar f_a^{-1}p_a f_a,~~g_a\rar g_af_a,~~f_a=f(z,\bz)|_{z=x_a}.
\eq
The corresponding vector fields are generated by the flatness constraint
\beq{2.4}
\bp A-\ka\p\bAp+[\bAp,A]-2\pi i\sum_{a=1}^np_a\de(x_a)=0,
\eq
where $\de(x_a)$ is a distribution defined by the integral
$$
\frac{1}{2\pi i}\int_{\Si_{g,n}}\bp\frac{1}{z-x_a},
$$
$z$ is a local coordinate in a neighborhood of $x_a$.

\bigskip
{\bf 2.Hamiltonians and times.}
The complex structure on $\Si_{g,n}$, defined locally by the $\bp$ operator,
 can be deformed by the Beltramy differential $\mu\in\Om^{(-1,1)}(\Si_{g,n})$:
$$
\bp\rar\bp+\mu\p,~~\mu=\mu(z,\bz)\frac{\p}{\p z}\otimes d\bz.
$$
It should be mentioned that $\mu$ depends on the choice of local coordinates as
$$
\mu(w,\bar{w})=
\frac{-\mu(z,\bz)\p w+\bp w}{\mu(z,\bz)\p \bar{w}-\bp \bar{w}},~~
w=w(z,\bz).
$$
In particular, if
$$
w=z-\ep(z,\bz),~~
\bar{w}=\bz
$$
and $\ep(z,\bz)$ is small, then
$$
\mu=\bp\ep(z,\bz).
$$
If $w(z,\bz)$ is a global diffeomorphism  than $\mu(w,\bar{w})$ is equivalent to
$\mu(z,\bz)$. The equivalence relations in $\Om^{(-1,1)}(\Si_{g,n})$ is the
moduli space of complex structures on $\Si_{g,n}$. The tangent space to the
moduli space is the Teichm\"{u}ller space ${\cal T}_{g,n}\sim
H^1(\Si_{g,n},\G)$, where $\G\in T\Si_g$.
From the Riemann-Roch theorem one has
\beq{2.6}
l=\dim {\cal T}_{g,n}=3(g-1)+n.
\eq
Let $(\mu_1^0,\ldots,\mu_l^0)$ be the basis in the vector space
$H^1(\Si_{g,n},\G)$.
We use the same notation $\mu$ for the elements from $H^1(\Si_{g,n},\G)$. Then
\beq{2.7}
\mu=\sum_{s=1}^lt_s\mu_s^0.
\eq

In particular, for the moduli related to the marked points one can
choose a
basis in the following form. Let $(z,\bz)$ be the local coordinates in
a neighborhood
of the marked point $x_a$ and ${\cal U}_a$ is a neighborhood of $x_a$ such
that $x_b\not\in{\cal U}_a$ if $x_b\neq x_a$. Define a $C^\infty$ function
\beq{chi}
\chi_a(z,\bz)=\left\{
\begin{array}{cl}
1,& z\in {\cal U}_a'\subset {\cal U}_a\\
0,& z\not\in {\cal U}_a.
\end{array}
\right.
\eq
Then in (\ref{2.7})
\beq{mp}
\mu_a=t_a\bp\chi_a(z,\bz),~~~
(\mu_a^0=\bp\chi_a(z,\bz),~ t_a=x_a^0-x_a).
\eq

We replace $\bAp$ by $\bA$
\beq{2.8}
\bA=\bAp+\f1{\ka}\mu A.
\eq
 The gauge transformations (\ref{2.3}) acts on $\bA$ as
\beq{2.9}
\bA\rar f^{-1}(\bp+\mu\p) f+f^{-1}\bA f.
\eq

The choice of complex structure defines the polarization $(A,\bA)$ of ${\cal R}^N$.
Thus, we define the bundle ${\cal P}_1^N$ over ${\cal T}_{g,n}$ with the local
coordinates
$$
(A,\bA,\bfp,\bft),~~\bfp=(p_1,\ldots,p_n),~\bft=(t_1,\ldots,t_l).
$$
$$
\begin{array}{cc}
{\cal P}^N_1& \\
\downarrow& {\cal R}^N\\
{\cal T}_{g,n} &
\end{array}
$$
The bundle ${\cal P}^N_1$ plays the role the extended phase space, while ${\cal R}^N$
is the standard phase space with non degenerate form (\ref{2.2}) and $t_1,\ldots,t_l$
are the  times.
Substitute $\bAp=\bA-\f1{\ka}\mu A$ in the symplectic form (\ref{2.2}). Then
it takes the standard form on the extended phase space \cite{Ar}
\beq{2.11}
\om=\om_0-\f1{\ka}\sum_{s=1}^ldH_sdt_s,
\eq
where
\beq{2.12}
\om_0=\int_{\Si_{g,n}}<dA~d\bA>+\om_{\cal O},
\eq
and
\beq{2.13}
H_s=\oh\int_{\Si_{g,n}}<A^2>\mu_s^0.
\eq
The symplectic form $\om$ is  defined on the total space of ${\cal P}_1^N$.
It is degenerate on $l$ vector fields $D_s$:~ $\om(D_s,\cdot)=0$, where
$$
D_s=\p_{t_s}+\f1{\ka}\int_{\Si_{g,n}}<A\frac{\de}{\de\bA}>\mu^0_s=
\p_{t_s}+\f1{\ka}\{H_s,\cdot\}_{\om_0},
$$
and the Poisson brackets are inverse to the non-degenerate form
$\om_0$ on the fibers.
 The vector fields $D_s$ define the
equations of motion for any function $f$ on ${\cal P}^N_1$
$$
\frac{df}{dt_s}=\p_{t_s}f +\f1{\ka}\{H_s,f\}_{\om_0}.
$$
In addition, there are the consistency conditions for the Hamiltonians
( the Whitham equations \cite{Kr})
\beq{WE}
\ka\p_sH_r-\ka\p_rH_s+\{H_r,H_s\}=0.
\eq
 Since the Hamiltonians (\ref{2.13}) commute there exists
 the tau-function
$$
H_s=\frac{\p}{\p t_s}\log \tau, ~~
\tau=\exp\oh\sum_{s=1}^l\int_{\Si_{g,n}}<A^2>\mu_s.
$$

In particular,
\beq{2.14}
\p_sA=0,~~~\p_s\bA=\f1{\ka}A\mu^0_s, ~~~\p_sp_a=0
\eq
and therefore
$$
A=A_0,~~\bA=\bA_0+\f1{\ka}\mu A_0.
$$

Let $\Psi\in \Om^{(0)}(\SR,{\rm Aut} P)$ be a solution of the linear system
\beq{2.15}
(\ka\p+A)\Psi=0,
\eq
\beq{2.16}
(\bp+\sum_{s=1}^lt_s\mu^0_s\p+\bA)\Psi=0,
\eq
\beq{2.17}
\ka\p_s\Psi=0,~~(\p_s=\p_{t_s}).
\eq
The monodromy of $\Psi$ is  the transformation
$$
\Psi\rar\Psi {\cal Y}, ~~{\cal Y}\in {\rm Rep}(\pi_1(\SR)\to\SL{N}).
$$
The equation (\ref{2.17}) means that the monodromy is independent on the times.
The equations of motion (\ref{2.14}) for $A$ and $\bA$ are
the consistency conditions (\ref{2.15}),(\ref{2.17}) and (\ref{2.16}),(\ref{2.17})
correspondingly. The consistency condition of (\ref{2.15}) and (\ref{2.16}) is
the flatness constraint (\ref{2.4}).

\bigskip
{\bf 3. Symplectic reduction}.
Up to now the equations of motion, the linear problem, and the tau function are
trivial.
The meaningful equations arise after the gauge fixing
with respect to (\ref{2.3}),(\ref{2.3a}) and imposing the corresponding constraints (\ref{2.4}).
 The set ${\cal R}_{red}^N$ of the gauge orbits
on the constraint surface (\ref{2.4}) is the moduli space of flat connections
$$
{\cal R}_{red}^N={\rm (\ref{2.4})}/{\cal G}_N={\cal R}^N//{\cal G}_N.
$$
Let us fix $\bA$:
\beq{2.30}
\bL=f^{-1}(\bp+\mu\p) f+f^{-1}\bA f.
\eq
For  stable bundles $\bL$ can be choose in a such way, that
$\bL=0$ for $g=0$, $\bL=$diagonal constant matrix for $g=1$, and at least
antiholomorphic for  $g>1$.
Then the dual field
\beq{2.31}
L=f^{-1}\ka\p f+f^{-1}Af
\eq
can be found from the moment equation (\ref{2.4})
\beq{2.18}
(\bp+\p\mu)L-\ka\p\bL+[\bL,L]=2\pi i\sum_{a=1}^np_a\de(x_a).
\eq
Here we preserve the notion $p_a$ for the gauge transformed element of ${\cal O}_a$.
The gauge fixing (\ref{2.30}) and the moment constraint (\ref{2.18})
kill almost all degrees of freedom. The fibers
${\cal R}^{red}=\{L,\bL,{\bf p}\}$ become finite-dimensional,
as well as the bundle ${\cal P}^{red,N}_1$:
\beq{2.18d}
\dim{\cal R}^N_{red}=2(N^2-1)(g-1)+N(N-1)n,
\eq
$$
\dim{\cal P}_{1~red}^N=(2N^2+1)(g-1)+(N^2-N+1)n.
$$
On ${\cal P}_{1~red}^N$ $\om$ (\ref{2.11})  preserves its form
\beq{red}
\om_0=\int_{\Si_{g,n}}<dL~d\bL>+
\om_{\cal O},~~H_s=H_s(L)=\oh\int_{\Si_{g,n}}<L^2>\mu_s^0.
\eq
But now, due to (\ref{2.18}), the system is no long free because
$L$ depends on $\bL$, ${\bf p}$. Moreover, because $L$ depends explicitely
on $\bft$, the system (\ref{red}) is non autonomous.

Let $M_s=\p_sff^{-1}$. Then the equations of motion on ${\cal R}_N$ (\ref{2.14})
take the form
\beq{2.19}
\ka\p_sL-\ka\p M_s+[M_s,L]=0,~~s=1,\ldots,l,
\eq
\beq{2.20}
\ka\p_s\bL-(\bp+\p\mu)M_s+[M_s,\bL]=L\mu_s^0.
\eq
The equations (\ref{2.19}) are the analog of the Lax equations. The essential
difference is the differentiation $\p$ with respect to the spectral parameter.
 The equation (\ref{2.20}) allows define $M_s$.
These equations reproduce
the Schlesinger system, Elliptic Schlesinger system, multi-component
generalization of the Painlev\'{e} VI equation \cite{LO1}.
The equations (\ref{2.19}), (\ref{2.20}) along
with (\ref{2.18}) are consistency conditions for the linear system
\beq{4.36}
(\ka\p+L)\Psi=0,
\eq
\beq{4.37}
(\bp+\sum_st_s\mu_s^0\p+\bL)\Psi=0,
\eq
\beq{4.38}
(\ka\p_s+M_s)\Psi=0,~~(s=1,\ldots,l_2).
\eq
The equations (\ref{4.38}) provides the isomonodromy
property of the system (\ref{4.36}), (\ref{4.37}) with
respect to variations of the times $t_s$.
For this reason we call the nonlinear equations (\ref{2.19})
 the Hierarchy of the Isomonodromic Deformations.

\bigskip
{\bf 4.Scaling limit.}
Consider the limit $\ka\to 0$. The value $\ka=0$ is called critical.
The symplectic form $\om$ (\ref{2.11})
is singular in this limit.
Let us  replace  the times
$$
t_s\rar t_s^0+\ka t^H_s,~~~(t^H_s-~{\rm Hitchin~times})
$$
and assume that the times $t_s^0, ~(s=1,\dots,l)$ are fixed.
After this rescaling the form  (\ref{2.11}) becomes regular.
The rescaling procedure means that we blow up a vicinity
 of the fixed point $\mu(0)=t_s^0\mu_s^{(0)}$
 in ${\cal T}_{g,n}$ and the whole dynamic
 is developed in this vicinity.
This fixed point is defined by the complex coordinates
\beq{fp}
w_0=z-\sum_st^0_s\ep_s(z,\bz),~~\bar{w}_0=\bz,~~
 \p_{\bar{w}_0}=\bp+\mu(0)\p.
\eq
For $\ka=0$ the connection $A$ is transformed into the one-form $\Phi$
(the Higgs field)
$
\ka\p +A\rar \Phi,
$
(see (\ref{2.3})).
Let
$
L^0=\lim_{\ka\to 0}L,~~\bL^0=\lim_{\ka\to 0}\bL.
$
Then we obtain the autonomous Hamiltonian systems with the form
$$
\om^H=\int_{\Si_{g,n}}<d L^0~d\bL^0>+
2\pi i\sum_{a=1}^n<d(p_ag_a^{-1})~d g_a>
$$
and the commuting time independent quadratic integrals
$$
H_s=\oh\int_{\SR}<L_0^2>\mu_s^0.
$$
 The phase space
${\cal R}_{red}^N$ turns into the cotangent bundle to the
moduli of stable holomorphic $\SL{N}$-bundles over $\Si_{g,n}$.

The corresponding set of linear equations has the following form.
The level $\ka$ can be considered as the Planck constant (see
(\ref{4.36})). We consider the quasi-classical regime
\beq{Ps}
\Psi=\phi\exp\frac{\cal S}{\ka},
\eq
where $\phi$ is a group-valued function and ${\cal S}$ is a scalar phase.
Assume that
\beq{me}
\frac{\p}{\p\bar w_0}{\cal S}=0,~~
\frac{\p}{\p t^H_s}{\cal S}=0.
\eq
In the quasi-classical limit we set
\beq{SW}
 \p_{w_0}{\cal S}=\la.
\eq
Define the Baker-Akhiezer function
$$
Y=\phi\exp\sum_{s=1}^lt_s^H\frac{\p{\cal S}}{\p t^0_s}.
$$
Then instead of (\ref{4.36}), (\ref{4.37}), (\ref{4.38}) we obtain in the
limit $\ka\to 0$
$$
(\la+L^0)Y=0,
$$
$$
(\bp_{{\bar w}_0}+\la\sum_{s=1}^lt_s\mu_s^0+\bL^0)Y=0,
$$
$$
(\p_s+M^0_s)Y=0,~~(s=1,\ldots,l).
$$
Note, that the consistency conditions for the first and the last
equations are the
standard Lax equations
\beq{2.30a}
\p_sL^0+[M_s^0,L^0]=0,
\eq
while the consistency conditions for the first and the second equations
is just the Hitchin equation, defining $L^0$
\beq{2.31a}
\p_{{\bar w}_0}L^0+[\bL^0,L^0]=2\pi i\sum_{a=1}^n\de(x_a)p_a.
\eq

It follows from (\ref{2.30a}) that the resulting Hamiltonian system is
completely integrable \cite{Hi,Ne}. The commuting integrals are
$$
H_{s,l}=\f1{j+1}\int_{\SR}<L_0^{j+1}>\mu_{s,j}^0,~~(j=1,\ldots,N-1).
$$
The differentials $\mu_{s,j}^0$ will be defined in next Section.

The gauge properties of the Higgs field allows to define the spectral
curve
$$
{\cal C}:~\det (\la +L)=0.
$$
The fixed times $t^0_s$ change the complex structure of the spectral curve.
The one-form $ \p_{w_0}{\cal S}$ (\ref{SW}) plays the role of the Seiberg-Witten differential on
${\cal C}$ (\ref{SW}). It is meromorphic on ${\cal C}$ in terms of the
deformed complex structure (\ref{me}).

\section {Flat bundles and higher times}
\setcounter{equation}{0}

{\bf 1. Higher times.}
The higher times $\mu_j$ are related to the differentials
$\Om^{(-j,1)}(\SR)$. They define generalized deformations of the operator
$\bp$ by adding the higher order operators $\p^j$ (see below (\ref{7.16}),(\ref{7.16a})).
 There is an equivalence relations provided by a groupoid action
on $\Om^{(-j,1)}(\SR)$ \cite{LO2}.
The space of orbits are the moduli space of generalized complex
structures ${\cal M}_{g,n}^{(j)}$.
The tangent space ${\cal T}_{g,n}^{(j)}(\SR)$ to ${\cal M}_{g,n}^{(j)}$ at $\mu_j=0$ is
isomorphic to $H^1(\SR,\G^{\otimes j})$. It has dimension
\beq{7.1}
l_j=\dim H^1(\SR,\G^{\otimes j})=(2j+1)(g-1)+jn,
\eq
($l_1=l$ (\ref{2.6})).
Let $(\mu_{1,j}^{(0)},\ldots,\mu_{l_j,j}^{(0)})$ be the basis of ${\cal T}_{g,n}^{(j)}$.
Locally
$$
\mu_{s,j}^{(0)}=\mu_{s,j}^{(0)}(z,\bz)\left(\frac{\p}{\p z}\right)^j\otimes d\bz.
$$
Then $\mu_j\in {\cal T}_{g,n}^{(j)}$ is represented as
\beq{7.2}
\mu_j=\sum_{s=1}^{l_j}t_{s,j}\mu_{s,j}^{(0)}.
\eq
The  parameters related to the marked points can be chosen in the form (see (\ref{mp}))
\beq{7.4}
\mu_a=\sum_{s=0}^{j-1}t_{s,a}(z-x_a)^s\bp\chi_a(z,\bz).
\eq
They cover ${\cal T}_{g,n}^{(j)}$ for the rational curves ${\bf C}P^1(n)$.
The whole set of times is  the linear space
$$
{\cal T}(k)_{g,n}=\oplus_{j=1}^k{\cal T}_{g,n}^{(j)}.
$$
For $k=1$ it is the Teichm\"{u}ller space defined above. From (\ref{7.1})
\beq{7.3}
l(k)=\dim{\cal T}(k)_{g,n}=((k+1)^2-1)(g-1)+\frac{k(k+1)}{2}n
\eq

\bigskip
{\bf 2. Extended phase space}.
We will use the same phase space ${\cal R}^N$ (\ref{2.1}) and $\om$ (\ref{2.2}).
To introduce the higher times consider an operator $\nabla^{(j,0)}$ acting from
$\Om^0(\Si_{g,n},E)\to\Om^{(j,0)}(\Si_{g,n},E)$. Locally, over a small disk it
takes the form
$$
\nabla^{(j,0)}=(\ka\p+A)^j.
$$
Then represent $\nabla^{(0,1)}\sim \bp+\bAp$ as
$$
\bp+\bAp=\bp+\bA-\f1{\ka}\sum_{j=1}^k\mu_j(\ka\p+A)^j.
$$
Here $\bA$ is a new field related to the initial fields $A,\bAp$ as
\beq{7.7}
\bA=\bAp+\f1{\ka}\sum_{j=1}^k\mu_j\ti{A}^{(j)}=
\bAp+\f1{\ka}\sum_{j=1}^k\ti{A}^{(j)}\sum_{s=1}^{l_j}t_{s,j}\mu_{s,j}^{(0)},
\eq
where $\ti{A}^{(j)}$ are defined by the recurrence relation
$$
\ti{A}^{(0)}=1,~~\ti{A}^{(j)}=(\ka\p+A)\ti{A}^{(j-1)}.
$$
Generically
\beq{7.5a}
\ti{A}^{(j)}=A^j+\ka\sum_{m=1}^{j-1}mA^{m-1}\p AA^{j-m-1}+
\ldots+(\ka\p)^{j-1}A.
\eq
In particular,
$$
 \ti{A}^{(1)}=A,~~~\ti{A}^{(2)}=A^2+\ka\p A,
$$
$$
\ti{A}^{(3)}=A^3+\ka(\p AA+2A\p A)+\ka^2\p^2A.
$$

The extended phase space ${\cal P}_k^N$ is a bundle over ${\cal T}(k)_{g,n}$
$$
\begin{array}{cc}
{\cal P}^N_k& \\
\downarrow& {\cal R}^N\\
{\cal T}(k)_{g,n} &
\end{array}
$$
 with the local coordinates
$$
(A,\bA,\bfp,\bft),~~
\bfp=(p_1,\ldots,p_n),~~\bft=(t_{s,j}),~s=1\ldots,l_j,~j=1,\ldots,k.
$$
The symplectic form $\om$ (\ref{2.2}) gives rise to the degenerate
symplectic form on ${\cal P}^N_k$.
In terms of $A$ and $\bA$  (\ref{2.2}) is equal
\beq{7.8}
\om=\int_{\Si_{g,n}}<dA~d\bA>+\om_{\cal O}-
\f1{\ka}\int_{\Si_{g,n}}
<dA~d(\sum_{j=1}^k\ti{A}^{(j)}\mu_{j})>.
\eq

For $k>1$ the form cannot be represented in the Hamiltonian form as before (\ref{2.11}).
Thus, the notion of the Whitham equations (\ref{WE}) and the tau-function become obscure.
On the other hand $\om$ is degenerate on the $l(k)$ vector fields
\beq{7.8d}
D_{(s,j)}=\ka\p_{t_{s,j}}+\mu_{s,j}^0\ti{A}^{(j)}\frac{\de}{\de\bA}.
\eq
The equations of motion for any observable $f(A,\bA,\bft)$ takes the form
$$
\p_{t_{s,j}}f(A,\bA,\bft)=D_{(s,j)}f(A,\bA,\bft).
$$
In particular, for $A$ and $\bA$ one has
\beq{7.12}
\p_{t_{s,j}}A=0,
\eq
\beq{7.13}
\p_{t_{s,j}}\bA=\f1{\ka}\mu_{s,j}^0\ti{A}^{(j)}.
\eq

It follows from (\ref{7.5a}) that the gauge action ${\cal G}_N$ on $A$ (\ref{2.3})
 induces the following transformations of $\ti{A}^{(j)}$
$$
\ti{A}^{(j)}\rar
f^{-1}
\left(\sum_{i=1}^j\ka^{j-i}C_j^i\ti{A}^{(i)}\p^{j-i}
\right)f,~~\left(C_j^i=\frac{j!}{i!(j-i)!}\right).
$$
Then for the new field $\bA$
\beq{7.10}
\bA\rar
f^{-1}
\left(\bp+\sum_{i=1}^k\mu_i\sum_{l=0}^{i-1}\ka^{i-l-1}C_i^l\ti{A}^{(l)}\p^{i-l}
\right)f+f^{-1}\bA f.
\eq
The constraints, generating these transformations, are read off from (\ref{2.4})
and (\ref{7.7})
\beq{7.18}
\bp A+\sum_{j=1}^k\p(\mu_j\ti{A}^{(j)})-\ka\p\bA+
[\bA-\f1{\ka}\sum_{j=1}^k\mu_j\ti{A}^{(j)},A]-2\pi i\sum_{a=1}^np_a\de(x_a)=0.
\eq
But, now the constraints become nonlinear (see (\ref{7.5a}))
As for the case $k=1$ the equations of motion (\ref{7.12}), (\ref{7.13})
 and the constraints
(\ref{7.18}) are the consistency conditions for the linear system (\ref{2.15}),
(\ref{2.16}) and (\ref{2.17}). In general case they take the form
\beq{7.15}
(\ka\p+A)\Psi=0,
\eq
\beq{7.16}
\left(\bp+
\f1{\ka}\sum_{j=1}^k\mu_j\sum_{m=0}^{j-1}C_j^m\ti{A}^{(m)}(\ka\p)^{j-m}+
\bA
\right)\Psi=0,
\eq
\beq{7.17}
\ka\p_{s,j}\Psi=0,~~(\p_s=\p_{t_{s,j}}).
\eq
The monodromy  of (\ref{7.15}) and (\ref{7.16}) is independent of the
whole set of times $\bft\in{\cal T}(k)$.

Consider for example the case $k=2$. The previous expressions take the form
$$
\om=\int_{\Si_{g,n}}<dA~(d\bA-\p dA\sum_{s=1}^{l_2}t_{s,2}\mu_{s,2}^{(0)})>
+\om_{\cal O}-\f1{2\ka}\int_{\Si_{g,n}}
\left (<dA^2>\sum_{s=1}^{l_1}\mu_{s,1}^{(0)}
\right)dt_{s,1}
$$
\beq{7.9}
-\f1{3\ka}\int_{\Si_{g,n}}
\left (<dA(A^2-\ka\p A)>\sum_{s=1}^{l_2}\mu_{s,2}^{(0)}
\right)dt_{s,2}.
\eq

The time evolutions of $\bA$ are defined as
\beq{7.14}
\p_{t_{s,1}}\bA=\f1{\ka}\mu^0_{s,1}A,~~
\p_{t_{s,2}}\bA=\f1{\ka}\mu^0_{s,2}(A^2+\ka\p A).
\eq
The gauge symmetries preserving $\om_0$
for $k=2$  have the form
\beq{7.11}
\bA\rar
f^{-1}
\left(\bp+\mu_1\p+\mu_2(\ka\p^2+2A\p)
\right)f+f^{-1}\bA f.
\eq
The constraints (\ref{7.18}) generating (\ref{7.11}) are
\beq{7.20}
\bp A+\p\left
(\mu_1A+\mu_2(A^2+\ka\p A)
\right)-\ka\p\bA+[\bA-\mu_2\p A,A]=\sum_{a=1}^np_a\de(x_a).
\eq
The linear equation (\ref{7.16}) takes the form
\beq{7.16a}
(\bp+
\mu_1\p+\mu_2(\ka\p^2+2A\p)+\bA)\Psi=0,
\eq
while (\ref{7.15}) and (\ref{7.17}) are the same. The operator in the left hand
side defined generalized holomorphic structure on the bundle $E$.

\bigskip
{\bf 3. Symplectic reduction}. As for the complex structure we use the
symplectic reduction to derive the nontrivial equations.
Let fix $\bA$ similar to (\ref{2.30})
\beq{7.21}
\bL=f^{-1}
\left(\bp+\sum_{i=1}^j\mu_i\sum_{l=0}^{i-1}\ka^{i-l-1}C_i^l\ti{A}^{(l)}\p^{i-l}
\right)f+f^{-1}\bA f,
\eq
where again $\bL=0$ for $g=0$, $\bL=$diagonal constant matrix for $g=1$,
and antiholomorphic for an arbitrary genus curves. Substituting $\bL$
and $L$ in (\ref{7.18}) one obtains
\beq{7.22}
\bp L+\sum_{j=1}^k\p(\mu_j\ti{L}^{(j)})-\ka\p\bL+
[\bL-\f1{\ka}\sum_{j=1}^k\mu_j\ti{L}^{(j)},L]
-\sum_{a=1}^np_a\de(x_a)=0,
\eq
where $\ti{L}^{(j)}$ is defined as $\ti{A}^{(j)}$
$$
\ti{L}^{(j)}=L^j+\ka+\sum_{m=1}^{j-1}mL^{m-1}\p LL^{j-m-1}
\ldots+\ka^{j-1}\p^{j-1}L.
$$
The symplectic reduction leads to the same phase space
${\cal R}^N_{red}=\{L,\bL,{\bf p}\}$. It defines the bundle ${\cal P}^{red,N}_k$ over
${\cal T}(k)_{g,n} $  and
$$
\dim{\cal P}^{red,N}_k=
\left(2(N^2-1)+(k+1)^2-1\right)(g-1)+(N^2-N+\oh k(k+1))n.
$$
The form $\om$ (\ref{7.8}) on   ${\cal P}^{red,N}_k$
$$
\om=\int_{\Si_{g,n}}<dL~d\bL>-
\f1{\ka}\int_{\Si_{g,n}}
<dL~d(\sum_{j=1}^k\ti{L}^{(j)}>\sum_{s=1}^{l_j}t_{s,j})\mu_{s,j}^{(0)}+\om_{\cal O}.
$$
is degenerate as on ${\cal P}^{N}_k$. But
now the equations of motion are no longer free. Nevertheless, after the
symplectic reduction (\ref{7.12}) and (\ref{7.13}) take the form of the Lax equations
\beq{7.23}
\p_{s,j}L-\ka\p M_{s,j}+[M_{s,j},L]=0,~~(s=1,\ldots,l_j;~j=1,\ldots,k),
\eq
\beq{7.24}
\bp M_{s,j}+[M_{s,j},\bL-
\f1{\ka}\sum_{j=1}^k\ti{L}^{(j)}\sum_{s=1}^{l_j}\mu_{s,j}]=
\ka\p_{s,j}\bL-
\sum_{j=1}^k\sum_{s=1}^{l_j}
(\ti{L}^{(j)}\mu_{s,j}^0+\mu_{s,j}^0t_{s,j}\p_{s,j}\ti{L}^{(j)})
\eq
As before, these equations along with the constraints (\ref{7.22})
are consistency conditions for the linear system
\beq{7.25}
(\ka\p+L)\Psi=0,
\eq
\beq{7.26}
(\bp+
\f1{\ka}\sum_{j=1}^k\mu_j\sum_{m=0}^{j-1}C_j^m\ti{L}^{(m)}(\ka\p)^{j-m}
+\bL)\Psi=0,
\eq
\beq{7.27}
(\ka\p_{s,j}+M_{s,j})\Psi=0,~~(\p_{s,j}=\p_{t_{s,j}}).
\eq

\section{ Integrable hierarchies in the scaling limit}
\setcounter{equation}{0}

In the limit  $\ka\to 0$ the Hamiltonian
form of equations is restored. We come to an autonomous hierarchy after the
same redifinition of the times
\beq{7.29}
t_{s,j}=t_{s,j}^0+\ka t^H_{s,j},
\eq
where $t_{s,j}^0$ are fixed and $t^H_{s,j}$ are  the genuine times.
We fix a point in the generalized Teichm\"{u}ller
space ${\cal T}(k)_{g,n}$
\beq{7.29a}
\mu_j(0)=\sum_{s=1}^{l_j}t_{s,j}^0\mu_{s,j}^0,~~(j=1,2\ldots,k)
\eq

\bigskip
{\bf 1.Scaling limit in the upstairs systems.}
Consider the "quasi-classical limit" of the linear problem
for the isomonodromy hierarchy (\ref{7.15}),(\ref{7.16}), and (\ref{7.17}).
Put as before (\ref{Ps})
$$
\Psi=\phi\exp\frac{\cal S}{\ka}.
$$
To come to the isospectral problem for $\Phi$ in
(\ref{7.15}) we define
\beq{7.30}
\p{\cal S}=\la.
 \eq
To exclude the singular terms of order $\ka^{-1}$ in (\ref{7.16}) and (\ref{7.17}) we
assume that the phase ${\cal S}$ satisfies in addition to (\ref{7.30})
\beq{mer}
(\p_{{\bar w}_0}-\sum_{j=2}^k(-1)^j\mu_j(0)\la^j){\cal S}=0,
\eq
$$
\frac{\p}{\p t^H_{s,j}}{\cal S}=0.
$$

Let $\Phi=\lim_{\ka\to 0} A$ be the Higgs field
$\Phi\in\Om^{(1,0)}(\Si_{g,n},\sll{N})$.
 As before, introduce the matrix Baker-Akhiezer function
$$
Y=\phi\exp\sum_{s,j}t_{s,j}^H\frac{\p{\cal S}}{\p
t_{s,j}^0}.
$$
Then in the limit $\ka\to 0$ (\ref{7.15}),(\ref{7.16}), and (\ref{7.17})
take the form of the isospectral problem
\beq{7.44b}
(\la+\Phi)Y=0,
\eq
\beq{7.45}
(\bp+\mu_1^0\p+\sum_{j=1}^k\mu_j(0)S_j(\Phi,\p\Phi)
-\sum_{j=1}^k\sum_{s=1}^{l_j}t_{s,j}^H\mu_{s,j}^0\Phi^j+\bA)Y=0,
\eq
\beq{7.46}
\p_{s,j}Y=0.
\eq
Here the field $S_j$ is a  polynomial in $\Phi$ with a linear dependence on $\p\Phi$.
It is defined as a result  of the action of the differential
operator in (\ref{7.16}) on $\Psi$ (\ref{Ps}) in the limit $\ka\to 0$
\beq{7.40}
S_j(\Phi,\p\Phi)=\lim_{\ka\to 0}
\f1{\ka}
\sum_{s=1}^{l_j}(t_{s,j}^0+t_{s,j}^H)
\sum_{m=0}^{j-1}C_j^m\ti{A}^{(m)}(\ka\p)^{j-m}
\left(\phi\exp\frac{\cal S}{\ka}\right).
\eq
The explicit form $S_j(\Phi,\p\Phi)$ for small $j$ is
\beq{7.47}
S_1(\Phi,\p\Phi)=0,~~S_2(\Phi,\p\Phi)=-2\p\Phi,~~
S_3(\Phi,,\p\Phi)=3\Phi\p\Phi,
\eq
$$
S_4(\Phi,\p\Phi)=-4\Phi^2\p\Phi+\p\Phi\Phi^2+6\Phi\p\Phi\Phi.
$$

Consider the consistency conditions for (\ref{7.44b}),(\ref{7.46}) and
(\ref{7.45}).(\ref{7.46})
\beq{7.48}
\p_{s,j}\Phi=0,~~\p_{s,j}\bA=\Phi^j\mu_{s,j}^0,~~
(\p_{s,j}=\frac{\p}{\p t^H_{s,j}}).
\eq
They can be interpreted as  the equations of motion of the Hamiltonian
system on
 ${\cal R}^N$  equipped by the  non degenerate symplectic form
$\om^H+\om_{\cal O}$
\beq{7.39}
\om^H=\int_{\Si_{g,n}}<d\Phi d\bAp>+\om_{\cal O},~~~
\bAp=\sum_{j=1}^k\mu_j(0)S_j(\Phi,\p\Phi)+\bA,
\eq
and the set of Hamiltonians
$$
H_{s,j}=\f1{j+1}\int_{\Si_{g,n}}<\Phi>^{j+1}t_{s,j}^0\mu_{s,j}^0,~~(j=1,\ldots,k).
$$
Then the equations of motion for this Hamiltonian hierarchy coincide
with  (\ref{7.48}) along with
$$
\p_{s,j} p_a=0,~~~~\p_{s,j}g_a=0.
$$
In particular, the Hamiltonian $H_{s,j}$ are in involution.

\bigskip
{\bf 2.Symplectic reduction.}
The gauge symmetries of this system is
\beq{7.41}
\Phi\rar f^{-1}\Phi f,~~
\bAp\rar f^{-1}\p_{{\bar w}_0} f+f^{-1}\bAp f.
\eq
The orbit degrees of freedom are transformed as before (\ref{2.3a}).
The moment constraints imposed by these symmetries take the form
\beq{7.44a}
\p_{\bar {w}_0} \Phi+
\sum_{j=2}^k\mu_j(0)[S_j(\Phi,\p\Phi),\Phi]+[\bA,\Phi]=2\pi i\sum_{a=1}^n\de(x_a)p_a.
\eq
For $k=2$, $\bAp=\bA-2\mu_2(0)\p\Phi$,
$$
\om^H=\int_{\Si_{g,n}}<d\Phi d\bA>-2\mu_2<d\Phi\p d\Phi>,
$$
the gauge transformations of $\bA$
$$
\bA\rar f^{-1}\left(
\bp_{{\bar w}_0}-2\mu_2(0)([\Phi,\p ff^{-1}]+\p\Phi)
\right) f+f^{-1}\bA f,
$$
and the moment constraints take the form
\beq{7.50}
\p_{{\bar w}_0}\Phi+[\bA-2\mu_2(0)\p\Phi,\Phi]=2\pi i\sum_{a=1}^n\de(x_a)p_a.
\eq

In general case, after the symplectic reduction we come
to the generalized Hitchin equation (GHE) depending on parameters
$\mu_j(0)\in{\cal T}(k)_{g,n}$
\beq{7.44}
\bp_{{\bar w}_0} L+
\sum_{j=2}^k\mu_j(0)[S_j(L,\p L),L]+[\bL,L]=2\pi i\sum_{a=1}^n\de(x_a)p_a,
\eq
where $\bL$ is assumed to be fixed by the gauge transformations (\ref{7.41})
and $L=f^{-1}\Phi f$.
The solutions of (\ref{7.44}) determine an autonomous Hamiltonian system
on the reduced phase space ${\cal R}_{red}^N=(L,\bL,\bfp)$. The symplectic form
on ${\cal R}_{red}^N$ is
\beq{7.52}
\om^H=\int_{\Si_{g,n}}
<\de L\de\bL'>+\om_{\cal O},~~~
\bL'=\bL-\sum_{j=2}^k\mu_j(0)S_j(L,\p L)
\eq
and
$$
H_{s,j}
=\f1{j+1}\int_{\Si_{g,n}}<L^{j+1}>t_{s,j}^0\mu^0_{s,j}.
$$
Since $H_{s,j}$ are in involution on ${\cal R}^N$ the same is true after
the symplectic reduction on ${\cal R}_{red}^N$.
 Assume now that $k+1=N$. Then the number of the integrals
is equal to
 $\dim{\cal T}(k)_{g,n}=\oh\dim{\cal R}^N$ (see (\ref{7.3}) and (\ref{2.18d})). Thereby,
the limiting autonomous system is completely integrable.
The fixed times $t^0_{s,j}$ can be considered as the deformation parameters of
the integrable system on the phase space ${\cal R}^N=(L_0,\bL_0,\bfp)$.

The accompanying linear problem takes the form
$$
(\la +L)Y=0,
$$
$$
(\bp_{{\bar w}_0}+\sum_{j=1}^k\mu_j(0)S_j(L,\p L)+\bL)Y=0,
$$
$$
(\p_{t^H_{s,j}}+M_{0,s,j})Y=0.
$$
The first equation allows to construct the spectral curve
$$
{\cal C}:~~\det(\la+L(w_0,\mu_2(0),\ldots,\mu_k(0))=0.
$$
The phase ${\cal S}$ in (\ref{Ps}) as before gives rise to  the Seiberg-Witten differential
(\ref{7.30}) on ${\cal C}$.  But in contrast with the case $k=1$ it
is no longer meromorphic in the complex structure $(w_0,\bar{w}_0)$ on
$\Si_{g,n}$ (\ref{mer}). The reason is that in general, if $k>1$,
then complex structures on
${\cal C}$ cannot be described in terms of complex structures on $\Si_{g,n}$.

\bigskip
{\bf 3.Solutions of GHE via perturbation expansions.}
GHE (\ref{7.44}) is a nonlinear differential matrix equation (for $k>1$).
The first nontrivial case is
$k=2$
$$
\p_{\bar {w}_0}L+[\bL-2\mu_2(0)\p L,L]=2\pi i\sum_{a=1}^n\de(x_a)p_a.
$$
It looks hopeless to find  explicit solutions of GHE as it was done for $k=1$ \cite{Ne}.
At the moment the only way
is to construct the series expansions assuming that the deformation parameters
$t^0_{s,j}$ are small. Therefore, we are investigating only a neighborhood
$\mu_j=0,~j=2,\ldots,k$ in the space of generalized complex structures
${\cal T}(k)_{g,n}$.

Consider deformations in a one specific direction $\mu^0_{s,j}$ and put
$\ve=t^0_{s,j}$. Let $L_0$ satisfies the usual Hitchin equation (\ref{2.31a}) and
represent $L$ as
$$
L=L_0+\sum_n\ve^nL_n.
$$
Substituting this expansion in GHE one obtain the chain
of recurrence differential relations
\beq{7.60}
\p_{\bar {w}}L_n+[\bL,L_n]=\mu_{s,j}^0[L_{n-1},S_j(L_{n-1},\p L_{n-1})].
\eq

Let $g=0$. Then, as we have already mentioned, $\bL=0$ and (\ref{7.60})
splits in the set of differential equations
$$
\p_{\bar {w}}(L_n)_{ml}=(F_{n-1})_{ml},~~(m,l=1,\ldots, N),
$$
where
$$
F_{n-1}=\mu_{s,j}^0[L_{n-1},S_j(L_{n-1},\p L_{n-1})],
$$
The solutions can be expressed in the integral form
\beq{7.61}
(L_n(w,\bar{w}))_{ml}=\int_{\Si_{g,n}}G(w-z,\bar{w}-\bz)(F_{n-1}(z,\bz))_{ml},
\eq
where $G$ is a Green function on a rational curve
$$
G=\f1{w-z},~~~\p_{\bar {w}}G=\de(z,\bz).
$$

Consider next the case $g=1,n=1$. For elliptic curves $\bL$ can be diagonalized\\
 $\bL=\di(u_1,\ldots,u_N)$
and thereby (\ref{7.60}) again  splits. Consider, for example, the non-diagonal
part of (\ref{7.60})
$$
\p_{\bar {w}}(L_n)_{ml}+(u_m-u_l)(L_n)_{ml}=
(F_{n-1})_{ml},~~(m\neq l,~m,l=1,\ldots, N).
$$
Then again we represent $(L_n)_{ml}$ as (\ref{7.61}), where
the Green function of this equation has the form \cite{Ne}
$$
G(z,\bz)=\exp(\frac{z-\bz}{\tau-\bar{\tau}}(u_m-u_l))
\frac{\te(u_m-u_l+z)\te'(0)}{\te(u_m-u_l)\te(z)}.
$$
In some  cases it is possible to calculate the integrals explicitly,
and thereby to find the
corrections to the Lax operator induced by small perturbations of
the generalized complex structures. The detail analysis of
corresponding Hamiltonian systems will be published elsewhere.

\small{

}
\end{document}